# Simulation and Instability Investigation of the Flow around a Cylinder between Two Parallel Walls

**Hua-Shu Dou\*, An-Qing Ben\***

Faculty of Mechanical Engineering and Automation, Zhejiang Sci-Tech University,, Hangzhou, Zhejiang 310018, China

The two-dimensional flows around a cylinder between two parallel walls at Re=40 and Re=100 are simulated with computational fluid dynamics (CFD). The governing equations are Navier-Stokes equations. They are discretized with finite volume method (FVM) and the solution is iterated with PISO Algorithm. Then, the calculating results are compared with the numerical results in literature, and good agreements are obtained. After that, the mechanism of the formation of Karman vortex street is investigated and the instability of the entire flow field is analyzed with the energy gradient theory. It is found that the two eddies attached at the rear of the cylinder have no effect on the flow instability for steady flow, i.e., they don't contribute to the formation of Karman vortex street. The formation of Karman vortex street originates from the combinations of the interaction of two shear layers at two lateral sides of the cylinder and the absolute instability in the cylinder wake. For the flow with Karman vortex street, the initial instability occurs at the region inner a vortex downstream of the wake and the center of a vortex firstly loses its stability inner a vortex. For pressure driven flow, it is confirmed that the inflection point on the time-averaged velocity profile leads to the instability. It is concluded that the energy gradient theory is potentially applicable to study the flow stability and to reveal the mechanism of turbulent transition.

**Keywords**：numerical simulation, cylinder, energy gradient theory, stability, inflection point

## Introduction

The flow around a cylinder and the wake behind the cylinder are classical issues of fluid mechanics [1]. This phenomenon is common in nature, such as the water flow past the bridge pier, the wind past a building and the air flow past the airfoil etc. It's also closely relevant to a large number of other practical applications, such as submarines, off shore structures, and pipelines etc. [2]. An important characteristic of this type of flow is the periodical vortex shedding in the cylinder wake at a sufficient high Reynolds number, and the wake can induce longitudinal and transverse unsteady loads acting on the object and excite vibrating response to the structure leading to seriously damage. Thus, the study on the mechanisms of flow separation and the shedding of vortexes is of great significance.

It has taken a long time in studies of the flow around a bluff body. Von Karmam firstly studied and analyzed the phenomenon, and he found the relationship between the structure of vortex and the drag acting on the cylinders in 1912. Later, a large number of researchers devoted to revealing the physic feature of the flow around the cylinders. Roshko [3] is the first one who found the existence of transition regime of the flow around a cylinder with experimental method, and he determined the existence of three different stages, i.e., linear flow, transition flow and disordered turbulence, in the cylinder wake between the low Reynolds and Middle Reynolds. Taneda [4-5] investigated the characteristic of eddies behind the cylinder by experiments. It was found that the Fopple eddies behind the cylinder occur when the Reynolds is 5, and the two eddies become longer and the frequency of vortex shedding accelerates as the Reynolds number increases. It was also found that Fopple eddies steadily attach on the rear of the cylinders when the Reynolds is smaller than 45, then the Karman vortex street is formed after the vortex shedding alternatively and moving downstream. He observed the formation of secondary vortex street after the vortex moving a long distance. Mathis [6] confirmed the existence of several modes proposed by Roshko [3], with experiment and in particular, he observed and interpreted a three-dimensional motion that occurs when two modes co-exist. Williamson [7] found with experiment that the three dimensional transition occurs, at Reynolds number 180-260. It is shown that there exist two modes related to three dimensional transitions near the cylinder wall.

---

\*Corresponding author : huashudou@yahoo.com
Hua-Shu Dou: Professor.

Zdravkovich [8] partitioned the flow into different modes according to the Reynolds, and they are as follows:

The flow begins to separate at lateral sides of the cylinder when 5<Re<40 and a pair of steady eddies occurs in the wake. The shedding vortex street forms at laminar state in the wake of the cylinder as 40<Re<200. The inner of the vortex begins to transit to turbulence when 200≤Re<300. The wake becomes fully turbulence as 300<Re<3×10$^5$, but the boundary layer remains as laminar flow.

The physical mechanism of the Karman vortex street is still not fully understood, and there exist four different descriptions about the formation of the vortex street:

Gerrard [9] described the mechanism of vortex shedding and the formation of vortex street, and he found that the determined factor leading to vortex shedding is the interaction of two separating shear layer at the lateral sides of the cylinder, i.e., interacting mode of shear layers. Coutanceau [10] meticulously investigated the formation of vortex with the secondary vortex oscillation mode by experiment, i.e. secondary vortex oscillation mode. Perry [11] found that once the vortex-shedding process begins, a so-called 'closed' cavity becomes open, and instantaneous 'alleyways' of fluid are formed which penetrate the cavity, i.e., open mode of the wake eddy. Trianiafyliou [12] Monkewitz and Nguyen [13] and Ortel [14] believed that the origin of vortex street is from the absolutely instability in the near wake, i.e., absolute instability mode.

The understandings on the mechanism of the formation of vortex street and the stability of flow field are deepened and promoted undoubtedly with all these studies. However, there still exist many problems to be resolved. For example, the reason why the center of a vortex firstly loses its stability and transits to turbulence in the cylinder wake is not known. The mechanism of the formation of vortex street is still controversial as discussed above. For example, Shi [15] pointed out that the instability in the cylinder wake originates from two inflection points at rear of the cylinder, but it doesn't always right for parallel shear driven flow, e.g. there always exists inflection point for plane Poiseuille-Couette flow, but the flow is stable when Re<2000.

With the aim to clarify these problems, the parameter distributions in the entire flow field for Re=40 and Re=100 are simulated with CFD in this paper. Then, the calculating results are compared with numerical results of Zovatto [16]. At last, the stability of the flow field is analyzed and the mechanism of the formation of Karman vortex street is investigated with the energy gradient theory.

**Briefly Introduction of the Energy Gradient Theory**

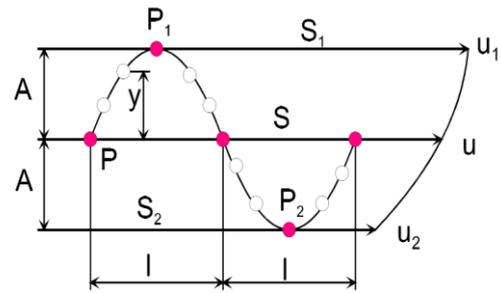

**Fig.1** Movement of a particle around its original equilibrium position in a cycle of disturbance

Dou et al. [17-20] proposed a new theory—the energy gradient theory, which is based on Newtonian mechanics and compatible with Navier-Stokes equations, to study the turbulent transition and the flow stability. The theory has been used to determine the flow stability and turbulent transition and good agreement with experiments has been obtained. The theory describes that: for a giving base parallel flow, the fluid particle moves ahead with oscillation when it is subjected to disturbance (see Fig.1). The fluid particle gains energy $\Delta E$ leading to amplification of a disturbance, and it also loses energy $\Delta H$ in streamwise direction which tends to absorb this disturbance and to keep the original laminar flow. The transition to turbulence depends on the relative magnitude of the two roles of energy gradient amplification and viscous friction damping under given disturbance. When the ratio reaches a critical value, the flow instability may be exited. So the determining criterion of instability can be written as follows:

$$F = \frac{\Delta E}{\Delta H} = \left(\frac{\partial E}{\partial n}\frac{2\bar{A}}{\pi}\right) \bigg/ \left(\frac{\partial H}{\partial s}\frac{\pi}{w_d}u\right) = \frac{2}{\pi^2}K\frac{A\omega_d}{u}$$

$$= \frac{2}{\pi^2} K \frac{v'_m}{u} < Const \quad (1)$$

$$K = \frac{\partial E/\partial n}{\partial H/\partial s} \quad (2)$$

Here, $F$ is a function of coordinates which expresses the ratio of the energy gained in a half-period by the particle and the energy loss due to viscosity in the half-period; K is a dimensionless field variable and expresses the ratio of transversal energy gradient and the rate of the energy loss along the streamline; $E = p + \frac{1}{2}\rho U^2$ is the total mechanical energy per unit volumetric fluid; s is along the streamwise direction, and n is along the transversal direction. H is the energy loss per unit volumetric fluid, u is the steamwise velocity of main flow; $\bar{A}$ is the amplitude of the disturbance distance, $\omega_d$ is the frequency of disturbance, and $v'_m = \bar{A}\omega_d$ is the amplitude of the disturbance of velocity.

Two criterions are proposed for pressure driven flow and shear driven flow based on the theory, and especially the criterion to determine the flow stability of pressure driven flow is described as: *the necessary and sufficient condition for turbulent transition is the existence of an inflection point on the velocity profile in the averaged flow* [19].

Now, the theory has been successfully applied to Taylor-Couette flow, plane Couette flow, plane Poiseuille flow and pipe Poiseuille flow, and it is found that the results show good agreement with experiments. These results demonstrate that the critical value of $K_{max}$ at subcritical turbulent transition for wall bounded parallel flows including both pressure driven and shear driven flows is $K_c$ =370-389 [17]. It means that the flow transition won't occur when the dimensionless parameter $K_{max}$ is less than $K_c$ in the flow field, otherwise it depends on the disturbance. Position with $K_{max}$ in a flow field firstly loses its stability, and position with large value of K will lose its stability earlier than that with small value of K.

## Physical Model and the Numerical Method

### Governing Equations and Numerical Method

In this paper, the flow is two-dimensional and keeps as laminar and the fluid used is water. Thus, the governing equations for incompressible fluid are as follows:

$$\nabla \cdot \vec{v} = 0 \quad (3)$$

$$\frac{\partial \vec{v}}{\partial t} + \vec{v} \cdot \nabla \vec{v} = -\frac{1}{\rho}\nabla p + \upsilon \nabla^2 \vec{v} \quad (4)$$

Here, ρ is the density, $\vec{v}$ is the velocity director. The governing equations are discretized with finite volume method (FVM). The coupling of pressure and velocity is done using PISO algorithm. The Reynolds number based on the cylinder diameter is defined as Re=$UD/\nu$, and $U$ is the average velocity at the inlet. D is the diameter of the cylinder, $\nu$ is the kinematic viscosity.

### Geometric Model and the Meshing

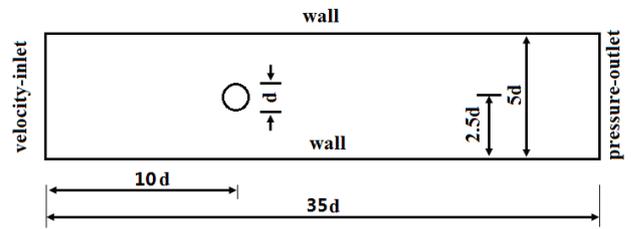

**Fig.2** Computational domain

The computational domain is shown in Fig.2. The flow field is modeled in two dimensional with the axes of the cylinder perpendicular to the direction of flow. The diameter of the cylinder is D=0.02m. The computational domain is 35D×5D. The upstream and downstream lengths are 10D and 25D from the center of the cylinder, respectively. The domain is partitioned into nine blocks, and the structured mesh is used for each block (see Fig.3).

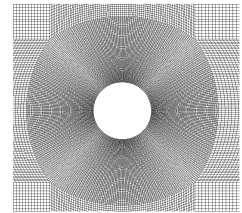

**Fig.3** The grid around the cylinder

### Characteristic Parameter of the Flow

The lift coefficient and drag coefficient are important characteristic parameters describing the fluid acting on the cylinder. And the Strouhal number is an important characteristic parameter describing the unsteady feature of vortex shedding. They are defined as follows:

$$C_l = \frac{F_l}{\frac{1}{2}\rho \bar{U}^2 A} \quad (5)$$

$$C_d = \frac{F_d}{\frac{1}{2}\rho \overline{U}^2 A} \quad (6)$$

$$S_t = \frac{fd}{\overline{U}} \quad (7)$$

Here, $F_l$ and $F_d$ represents the drag and the lift force respectively, A is the area projected in the flow direction, and the magnitude of A is determined by the diameter of the cylinder in two-dimensional flow, $\overline{U}$ is the average velocity, and $f$ is the frequency of vortex shedding.

### Boundary Conditions

As shown in Fig.2, the velocity inlet boundary condition is applied at the upstream of the cylinder. At the downstream, a pressure outlet boundary condition is defined. No-slip boundary condition is applied on the walls and the cylinder, i.e., u=0, v=0.

### Grid Independence Test

In this paper, the condition of Re=100 is taken as an example to examine the grid independence. Three meshes are used and they are marked as M1, M2 and M3 respectively. The minimum size of the three meshes is $\Delta X_{1min}=3.924\times10^{-4}$m, $\Delta X_{2min}=1.963\times10^{-4}$m and $\Delta X_{3min}=2.797\times10^{-5}$m respectively. The average drag coefficient $\overline{C_D}$ is selected as a reference to validate the calculating results. The simulation results of present study and the data in literature are shown in Fig.4. It can be found that good agreement is obtained with the experimental results. Finally, M2 is selected as the calculating grid.

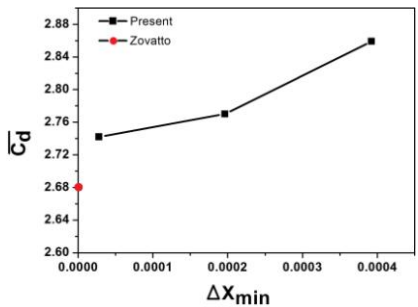

**Fig.4** Grid independence test

## Simulating Results and Analysis

### Comparison of the Calculating Results with those in Literature

The Strouhal number and the drag coefficient of present study and those in reference are shown in Table.1. It can be found that good agreement has been achieved with reference. As such, the calculating method used in this paper is reliable.

**Table.1** The results of present study and those in reference

|  | Strouhal Number | Drag Coefficient |
|---|---|---|
| Present | 0.2814 | 2.68 |
| Reference[20] | 0.2739 | 2.77 |
| Relative Error | 2.74% | 3.36% |

### Calculation of the Energy Gradient Function K

According to the energy gradient theory [18], the equation of the dimensionless parameter K can be written as:

$$K = \frac{\partial E/\partial n}{\partial H/\partial s} \quad (8)$$

Here the total mechanical energy E can be written as $E = P + \frac{1}{2}\rho U^2$, and the total velocity can be written as $U = \sqrt{u^2+v^2}$. Here $u$ is the velocity along the X direction, $v$ is the velocity along the Y direction, P is the static pressure of flow field, $n$ is the normal direction of the streamline, and s is the streamwise direction along the streamline.

For pressure driven flows, Eq.(8) can be written as

$$K = \frac{\partial E/\partial n}{\partial E/\partial s}$$

The energy gradient in the normal direction of a streamline can be written as [20]:

$$\frac{\partial E}{\partial n} = \frac{\partial(P+\frac{1}{2}\rho U^2)}{\partial n}$$

$$= \rho(\vec{U}\times\vec{\omega})\bullet\frac{d\vec{n}}{|d\vec{n}|} + (\mu\nabla^2\vec{U})\bullet\frac{d\vec{n}}{|d\vec{n}|} \quad (9)$$

The energy loss in the streamwise direction along a streamline for pressure driven flow can be written as [18][20]:

$$\frac{\partial E}{\partial s} = \frac{\partial(P+\frac{1}{2}\rho U^2)}{\partial s}$$

$$= \rho(\vec{U}\times\vec{\omega})\bullet\frac{d\vec{s}}{|d\vec{s}|} + (\mu\nabla^2\vec{U})\bullet\frac{d\vec{s}}{|d\vec{s}|} \quad (10)$$

Thus, the dimensionless parameter K of the energy gradient function can be written as:

$$K = \frac{\partial E/\partial n}{\partial E/\partial s}$$

$$= \frac{\rho(\vec{U}\times\vec{\omega})\bullet\dfrac{d\vec{n}}{|d\vec{n}|} + (\mu\nabla^2\vec{U})\bullet\dfrac{d\vec{n}}{|d\vec{n}|}}{\rho(\vec{U}\times\vec{\omega})\bullet\dfrac{d\vec{s}}{|d\vec{s}|} + (\mu\nabla^2\vec{U})\bullet\dfrac{d\vec{s}}{|d\vec{s}|}} \quad (11)$$

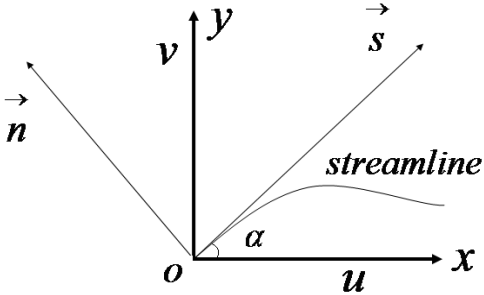

**Fig.5** Geometric relationship of physical qualities

As is shown in Fig.5, a streamline passes by point O, i.e., the origin of X axis and Y axis. From Fig.5, it can be obtained that:

$$\cos\alpha = \frac{u}{U}, \qquad \sin\alpha = \frac{v}{U} \quad (12)$$

$$\frac{d\vec{n}}{|d\vec{n}|} = (-\sin\alpha, \cos\alpha) \quad (13)$$

$$\frac{d\vec{s}}{|d\vec{s}|} = (\cos\alpha, \sin\alpha) \quad (14)$$

$$\omega_z = \frac{dv}{dx} - \frac{du}{dy}, \quad \vec{U}\times\vec{\omega} = (v\omega_z, u\omega_z) \quad (15)$$

Then the equation (11) can be written as:

$$K = \frac{\partial E/\partial n}{\partial E/\partial s}$$

$$= \frac{\mu(\dfrac{\partial^2 v}{\partial x^2}+\dfrac{\partial^2 v}{\partial y^2})\cos\alpha - \mu(\dfrac{\partial^2 u}{\partial x^2}+\dfrac{\partial^2 u}{\partial y^2})\sin\alpha}{\mu(\dfrac{\partial^2 v}{\partial x^2}+\dfrac{\partial^2 v}{\partial y^2})\sin\alpha + \mu(\dfrac{\partial^2 u}{\partial x^2}+\dfrac{\partial^2 u}{\partial y^2})\cos\alpha} -$$

$$\frac{-\rho v\omega_z \sin\alpha - \rho u\omega_z \cos\alpha}{\mu(\dfrac{\partial^2 v}{\partial x^2}+\dfrac{\partial^2 v}{\partial y^2})\sin\alpha + \mu(\dfrac{\partial^2 u}{\partial x^2}+\dfrac{\partial^2 u}{\partial y^2})\cos\alpha} \quad (16)$$

## Calculating Results and Analysis

### Flow at Re=40

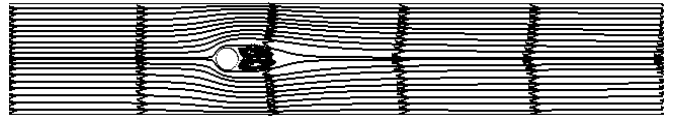

(a) Streamline

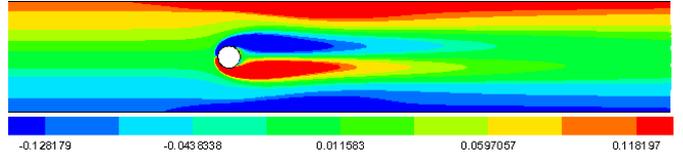

(b) Vorticity contour

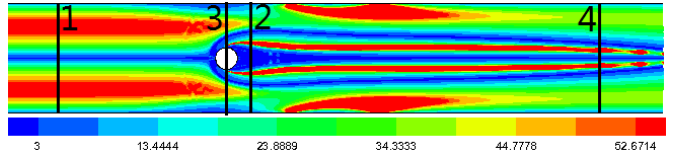

(c) K contour

**Fig.6** Distribution of Streamline, Vorticity and K at Re=40

As shown in Fig.6 (a) and (b), it can be found that there is no Karman vortex street in the cylinder wake and there are two eddies attached at the rear of the cylinder. The flow is laminar for plane-Poiseuille flow, when the Reynolds number is 40. The K contour of this case is shown in Fig.6 (c), and it can be found that the distribution of K ahead of the cylinder (inlet region) is in accordance with Dou's [18] equation $K = 0.75\cdot\text{Re}\cdot\dfrac{y}{h}\cdot(1-\dfrac{y^2}{h^2})$ (here h is half of the height between two walls). The detailed numerical distribution is shown in Fig.7. It can be seen from Fig.7 (b) that the magnitude of K is about 57, which is in accordance with theoretical value. The $K_{max}$ of this profile is located at $\dfrac{y}{h} = 0.58$, which is in accordance with experiments [21].

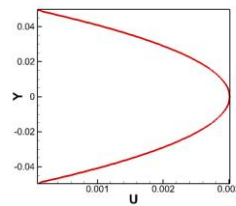 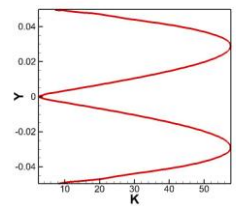

(a) Velocity  (b) K

**Fig.7** The distribution of velocity and K at section 1

Comparing the velocity contour with K contour, it can be found that the K at the position of the two eddies is very low. It represents that the two eddies at the rear of the cylinder have no effect on flow stability according to the energy gradient theory. To clarify this problem clearly, the detailed distributions of parameters at section 2 are shown in Fig.8. In Fig.8 (c), it can be found that the value of K at lateral sides of two eddies is large and the value of K at the position with two eddies is low, which numerically validates the conclusion that the two eddies at the rear of the cylinder have no effect on the flow stability. It can also be found that the $K_{max}$ at this section is located at the position where the second derivative of velocity is zero, i.e., the inflection point.

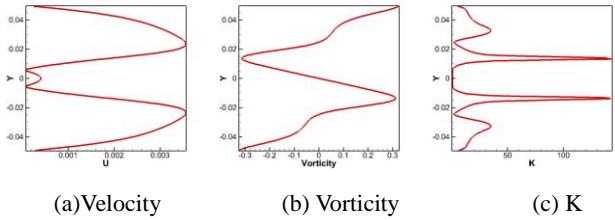

(a) Velocity     (b) Vorticity     (c) K

**Fig.8** Distribution of physical quantities at section 2

It can also be found that K at lateral sides of the cylinder is large. To study the mechanism, the detailed distributions of parameters at section 3 are shown in Fig.9. As mentioned above, $K_{max}$ at this section is also located at the position with inflection point in velocity profile.

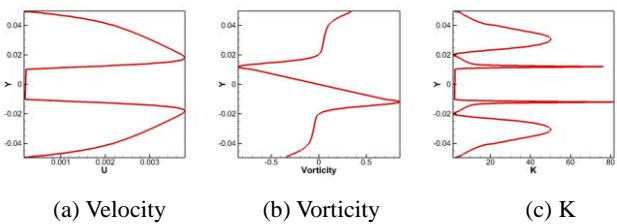

(a) Velocity     (b) Vorticity     (c) K

**Fig.9** Distribution of physical quantities at section 3

According to the energy gradient theory, the position with $K_{max}$ in the entire flow field will firstly lose its stability. At Re=40 condition, it can be found that the $K_{max}$ of the entire flow field occurs at section 4 (see Fig.6). To investigate the reason why the K here is largest, the detailed distributions of parameters at section 4 are shown in Fig.10. As shown in Fig.10, it can be found that $K_{max}$ occurs at the position with inflection point of velocity and maxima of vorticity. According to the definition of vorticity in two-dimensional flow, $\omega_z = \frac{\partial v}{\partial x} - \frac{\partial u}{\partial y}$. To get the maxima of vorticty, let $\frac{\partial \omega_z}{\partial y} = 0$, i.e., $\frac{\partial \omega_z}{\partial y} = \frac{\partial^2 v}{\partial x \partial y} - \frac{\partial^2 u}{\partial y^2} = 0$. For the flow around a cylinder between two parallel walls, $\frac{\partial^2 v}{\partial x \partial y}$ is very low. Thus, the equation $\frac{\partial \omega_z}{\partial y} = 0$ is proximately equal to $\frac{\partial^2 u}{\partial y^2} = 0$. Therefore, the maxima of the vorticity corresponds to the inflection point of velocity and the location of $K_{max}$ at each section.

The formation of Karman vortex street has been investigated by many researchers, but the mechanism is still not clear and controversial conclusions exist in literature. As mentioned above, it is found that the $K_{max}$ is located at section 4. According to the concept of absolute instability [12] [13] [14], the instability here will spread to the upstream and affect the stability of the two shear layers which are located at lateral sides of two eddies and are with large K. Then, the interaction of two shear layers leads to vortex shedding and formation of the Karman vortex street. This is consistent with the conclusion obtained by Gerrard [9] that the determined factor leading to vortex shedding is the interaction of two separating shear layers at the lateral sides of the cylinder. In summary, vortex shedding is originated from the combination of the interaction of two shear layers and absolute instability in the cylinder wake, which leads to the formation of Karman vortex street.

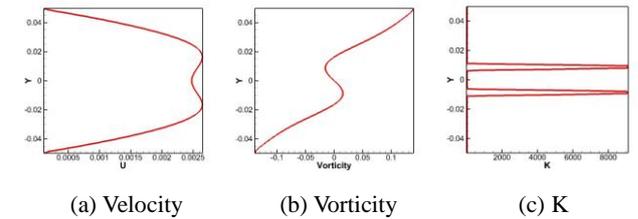

(a) Velocity     (b) Vorticity     (c) K

**Fig.10** Distribution of physical quantities at section 4

**Flow at Re=100**

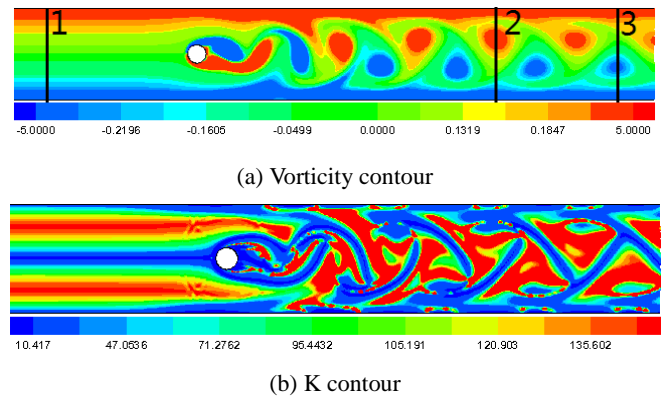

(a) Vorticity contour

(b) K contour

**Fig.11** Distributions of vorticity and K at Re=100

The flow at Re=100 is calculated and the vortcity contour and K contour are shown in Fig.11. As described above, the flow ahead of the cylinder and after the velocity inlet is laminar, and the distribution of K is in accordance with Dou's [18] equation mentioned above. The detailed distribution of K is shown in Fig.12. The magnitude of K is about 144, which is in accordance with theoretical value. The $K_{max}$ of this profile is located at $\frac{y}{h}=0.58$, which is also in accordance with experiments [21].

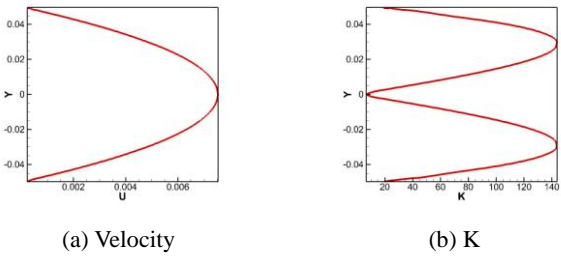

(a) Velocity          (b) K

**Fig. 12** Distribution K velocity and K at section 1

The stability of the cylinder wake is investigated with the energy gradient theory below. Through further investigation, it is found that $K_{max}$ of the entire flow field is located at each vortex in the cylinder wake. To study the mechanism, the detailed distributions of parameters at sections 1 and 2 are shown in Figs.13 and 14, respectively.

Comparing the velocity, vorticity and K contours in Fig.13 and Fig.14, it can also be found that the location of $K_{max}$ at each cross section corresponds to the inflection point of the velocity profile and the maxima of vorticity as mentioned above. It can also be found that, $K_{max}$ is located in a vortex center for each vortex, which represents that the flow will firstly lose its stability in a vortex center in the cylinder wake. In fact, the finding is in agreement with the conclusion in Zdravkovich [8]: the inner of the vortex begins to transit to turbulence when 200≤Re<300. In this part, the results validate and deepen the conclusion: the inner of the vortex will firstly lose its stability in entire flow field, and the center of a vortex will firstly lose its stability at each vortex. The reason is that it is the inflection point on the velocity profile that leads to the instability.

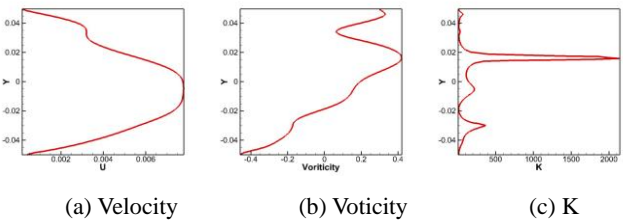

(a) Velocity      (b) Voticity      (c) K

**Fig.13** Distribution of physic quantities at section 2

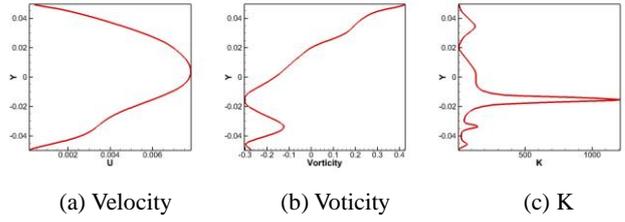

(a) Velocity      (b) Voticity      (c) K

**Fig.14** Distribution of physic quantities at section 3

## Conclusions

In this paper, the flows around a cylinder between two parallel walls at Re=40 and Re=100 are simulated with computational fluid mechanics (CFD). The simulating results are compared with the numerical results in literature and good agreement is obtained. Then the instabilities of flow field are investigated with the energy gradient theory. The formation of Karman vortex street is studied with Re=40. The initial instability of entire flow field with vortex street in the cylinder wake is investigated with Re=100. Several conclusions are obtained as follows:

1. The two eddies at the rear of the cylinder have no effect on the flow instability for steady flow, i.e., they don't contribute to the formation of Karman vortex street.
2. The formation of Karman vortex street originates from the combinations of the interaction of two shear layers at two lateral sides of the cylinder and the absolute instability in the cylinder wake.
3. For the flow with Karman vortex street, the initial instability occurs at the region inner a vortex and the center of a vortex firstly loses its stability inner a vortex.
4. For pressure driven flow, it is confirmed that the inflection point on the time-averaged velocity profile leads to the instability.
5. The energy gradient theory is potentially applicable to study the flow stability and it can profoundly reveal the mechanism of turbulent transition.

## Acknowledgement
This investigation is supported by the Special Major Project of Science and Technology of Zhejiang province (No. 2013C 01139), and the Science Foundation of Zhejiang Sci-Tech University (No. 11130032661215).